\documentclass{article}
\begin{document}
\title{Violation of The Unitarity Bound in a Finite $N=8$ Theory of Super Gravity}
\author{Ivan J. Muzinich \\
IJM Mathematical Consulting \\
1548 Carol Street \\
Camano Island WA 98282 \\
ijm@wavecable.com
}
\date{April 2011 }
\maketitle

\begin{abstract}

We comment on recent results of a possible finite theory of Super Gravity from both Feynman graph and global $E_{7(7)}$ symmetry arguments. The four point amplitude can be written as a series in the gravitational coupling and energy squared,$Gs$, multiplied by coefficient functions of the scattering angle. Every order of the perturbation expansion violates the unitarity bound for any element of the $S$ matrix, $|S|\leq1$ by powers for
large values of the energy beyond the Planck mass. While this conclusion is not unexpected; the meaning of such a theory is not understood. Either the theory is not finite or a non-perturbative completion is necessary. We first review the semi-classical eikonal methods, and then examine Borel techniques to sum such a series. Although many details are unknown, We present an example of how an asymptotic behavior that is power bounded and consistent with unitarity might emerge.

\end{abstract}

\pagebreak

Historically there has been no shortage of communications stressing the non-renormalizability of point like theories of perturbative quantum gravity based upon the Einstein Hilbert action with its incumbent coupling constant of length squared. Known works by Feynman[1], Mandelstam[2], Dewitt[3], t'Hooft and Veltman[4], Goroff, and Sagnotti[5], van deVen[6] and many others are of particular importance. However, in the distant past the emergence of a gauged super symmetry popularly known as super gravity piqued the theoretical physics community to embrace a finite theory of super gravity to two loop order and perhaps beyond. One loop is finite on shell because the only available counter term,$R^2$, is  a total divergence. Two loops are finite because of the lack of super symmetry of the possible $R^3$ counter term. The counter terms are given by the appropriate contractions of the Riemann tensor. See also the work of Deser, van Nieuwenhuizen, and collaborators[7][8].

\vspace{.25in}

In addition the past two decades have seen significant and beautiful calculations by many authors Bern, Carrasco, Dixon, Johansson, Roiban, Rosower and many collaborators[9][10][11][12] made  the case for a finite theory of $N=8$ super gravity based upon  Feynman graphs with $S$ matrix techniques. This work computes the discontinuities or maximal cuts of graphs to four loops. The maximal cuts relate the amplitudes to tree graphs which are products of the tree graphs for super symmetric $N=4$ Gauge Theory, SYM. The maximal cut graphs differ from the complete Feynman graphs by polynomials in the loop momenta which are alleged to disappear with the absence of bubble and triangle graphs. Recently a proof of finiteness to all orders has been announced by Kallosh[13] based upon a global non-compact $E_{7(7)}$ invariance of the S-matrix. While this work is very important, the proof is conditioned on the absence of anamolies in the conservation of Noether $E_{7(7)}$ currents.

\vspace{.25in}

The main thrust of this short communication is physical. We re-emphasize that unitarity is violated by powers of the energy by known results from Feynman graphs through at least four loops and to any finite order. The issue is the growth with energy of the perturbative contributions.  Unitarity is an efficient tool to state the argument. Such results were pointed out long ago for lower order results. This is probably the crispest contradiction between quantum mechanics and perturbative general relativity formulated on a flat Lorentz invariant background. Unitarity apparently is not a necessary condition for finiteness, although it is closely related to renormalizability in other field theories. There remains the question:
\vspace{.25in}

{\em What is the meaning of a finite quantum field theory that violates the unitary bound?}

\vspace{.25in}
\pagebreak
Therefore a non-perturbative completion of the theory becomes important. Super String or M Theory on the other hand has the advantage that the fixed angle scattering is exponentially damped at high energy to any finite loop order as emphasized by Gross and Mende[14] and others long ago.

\vspace{.25in}

Firstly, the known results to L loop order from Feynman graphs can be summarized by power counting, explicit calculation, and dimensional arguments by the following expression for the MHV(maximally helicity violating amplitude):

\begin{equation}
A_L(s,t,u)=M_{tree}stu\sum_P\kappa^{2L}P_L(s,t)C_L(s,t),
\end{equation}
where $M_{tree}$ is an appropriately symmetrized tree amplitude, $P_L(s,t)$ is a polynomial of order $L-3$, and $\sum_P$ is a sum over permutations . The amplitudes $C_L(s,t)$ are analytic functions of the Mandelstam invariants s,t, and u with $S$ matrix (Landau Bjorken) type singularities and are power counting finite with degree of divergence $D=4+6/L$ identical to the SYM theory. Here, $\kappa^2 =32\pi G$. These results are known explicitly to four loop order and correspond to the absence of counter terms of the form  $D^{2k}R^4$ [15][16] and are true to all orders if the arguments based upon the $E_{7(7)}$ symmetry are correct.

\vspace{.25in}
We next review the unitarity bound on the $S$ matrix. The difficulties with unitarity effect the field theory amplitudes and finite momentum transfer string theory amplitudes as well. The unitarity condition takes the form:

\begin{equation}
S^{\dagger}S=I
\end{equation}

Of course these considerations have been traditionally applied to amplitudes of finite mass particles not the 256 massless particles of the $N=8$ super multiplet. It immediately follows that any element of the $S$ matrix is bounded by unity $|S|\leq1$. The elements of the $T$ matrix, $T=\frac{S-1}{2i}$  are bounded by unity as well, $|T|\leq1$.
\space
Such an $S$ matrix does not exist for amplitudes with any finite number of massless particles in $D=4$ due to infrared problems. Infrared singularities have sources due to soft and collinear emission. The collinear effects are suppressed for gravity amplitudes because of the tensor nature of the coupling to gravitons in the numerator as mentioned by Weinberg many years ago[17]. Soft emissions can be settled at the level of transition probabilities $P_{fi}$ given by

\begin{equation}
P_{fi}=\sum_{ab}\rho^a_{f}\rho^b_{i}T_{ab}\overline{T_{ab}}\leq1,
\end{equation}
where $\rho_{f}$ and $\rho_{i}$ are the density matrices for final and initial observers and the sum is over degenerate states in accordance with expectations based upon the Lee Nauenberg, Kinoshita theorems[18][19].

\vspace{.25in}

It is clear that the energy growth of the MHV amplitude violates the unitarity bounds of the partial wave $S$ matrix and transition probabilities Equ.[3]. The violation of unitarity in the field theory amplitudes comes from fixed angle as well as fixed t which is in contrast to string theory where unitarity violation is limited to fixed t. One of two possible conclusions are possible:

\begin{itemize}
\item Either the field theory is not finite and another ultraviolet completion has to be found such as string theory. However, string theory finiteness does not necessarily imply field theory finiteness due to the lack of decoupling from the infinite tower of massive states.

\item Or the theory is finite and is at best an effective field theory for energies below the Planck scale. If we insist on a theory smooth through the Planck scale, then a non-perurbative completion has to be implemented to solve the unitarity issue.

\end{itemize}
\vspace{.25in}
It has been known for at least two decades that an expansion of the amplitude at large distances with the semi-classical eikonal representation is successful in restoring unitarity [20][21].  In the following we use Planck units for energies and distances, $G=M_{Pl}^{-2}$, where $M_{Pl}$ is the Planck mass. For example the lowest order single graviton exchange,

\begin{equation}
A=\frac{s^2}{t}
\end{equation}
violates unitarity in the above sense. However, at large impact parameter $b$, the amplitude can be resumed into the two dimensional Fourier transform valid in $D=4$:

\begin{equation}
A=s\int exp{i}{\vec{b}\cdot\vec{q}}[exp-iF(s,b)- 1] d^{2}b.
\end{equation}
The eikonal function $F(s,b)$ has an expansion in powers of $s^\frac{1}{2}/b$ at large distances:

\begin{equation}
F=s(lnb)+\sum_p a^p(s^\frac{1}{2}/b)^p.
\end{equation}
The $lnb$ comes from the transverse Fourier transform of the lowest order graviton exchange. The corrections in powers of, $s^\frac{1}{2}/b$ have been considered for elastic scattering and absorption by Amati, Ciafaloni, and Veneziano in reference [20]. While these considerations are valid for restoring unitarity at large energies and fixed momentum transfers, large distances, they are not suitable for the fixed angle small distance high energy behavior. Therefore, other techniques have to be explored.

\pagebreak

We now turn to Borel techniques[22]to explore the sum of a perturbation series of the type in Equ.[1]. We really do not know the growth of the perturbation series in $L$ loop order or if these techniques can be applied to any theory of gravity. If gauge theories are a guide, the number of contributions grow like at least like $L!$ in each loop order as remarked by Dyson for Quantum Electrodynamics many decades ago[23]. Therefore,from Equ.[1], we expect a series of the form:

\begin{equation}
A(s,\theta)=s\sum_L L!(s)^L f_L(\theta)
\end{equation}
with largely unknown coefficient functions  $f_L(\theta)$ of the fixed non-zero scattering angle with zero radius of convergence.  The Borel transform is

\begin{equation}
 B(t,\theta)=\sum_L t^L f_L(\theta)
\end{equation}
From Equ.[7], the original amplitude can be constructed with the Borel transform

\begin{equation}
A(s,\theta)=s\int_{0}^{\infty} due^{-u}B(su,\theta).
\end{equation}

As an illustrative example, if $f_L$ is assumed to be a constant in Equ.[7], then for large orders one obtains a simple dispersion relation for the amplitude:

\begin{equation}
A(s)=\int_{0}^{\infty} du\frac{e^{-u}}{u-s^{-1}\pm i\epsilon}.
\end{equation}
Equ.[10] defines an analytic function of $s$ almost everywhere with a branch point at $s=\infty$ where the Borel representation becomes singular. One has to specify the contour indicated by the $\pm i\epsilon$ prescription. For physical region Landau Bjorken $S$ matrix singularities [24][25][26], the physical Riemann sheet is approached from above $+i\epsilon$; however, the singularities in Equ.[10] are in both energy and coupling and are not defined by the usual $S$ matrix prescription.

\vspace{.25in}

The transform Equ. [10] is well defined for $s<0$ alternating sign series but not well defined with an imaginary part of order $\pm e^{-\frac{1}{s}}$ for $s\geq0$ same sign series. This is a signal for non-perturbative renormalon effects associated with Landau ghost type singularities noticed in applications of the renormalization group. This is often referred to as the Borel ambiguity. In any case an optimistic outcome of Equ.[10] would give a well behaved logarithmic asymptotic behavior, since the ambiguity effects the imaginary part in a negligible manner at strong coupling high $s$:

\begin{equation}
A(s)=O(lns),{s\rightarrow\infty}
\end{equation}
for large values of $s$. An interesting line of research would be the investigation of the combinatoric growth of the perturbation series for $N=4$ SYM, in particular its implications for $N=8$ super gravity. I wish re-emphasize that these arguments do not have the status of rigor and may provide at best a guide to the truth.

\vspace{.25in}
We wish to thank many colleagues for discussions namely Geoffrey Chew, Stanley Mandelstam, Joe Polchinski, Frank Paige, Marshall Baker, Larry Yaffe, Dan Freedman, and lastly but most importantly Henry Lubatti for encouragement to write this communication.

\vspace{.25in}

The author was formerly affiliated in his working career with Brookhaven National Laboratory, Upton NY and The Kavli Institute of Theoretical Physics, University of California, Santa Barbara CA. The author has been a consultant to the Honeywell Corporation, Redmond WA and Winsor Lighting Corporation, Seattle WA.
\vspace{.25in}

\begin{enumerate}
\item R.P. Feynman, Acta Phys.Polon, 24,697,(1963).
\item S. Mandelstam, Phys.Rev. 175,1580,1604,(1968).
\item B.S. Dewitt, Phys. Rev. 162,1195,1239,(1967).
\item G. t'Hooft and M. Veltman, Annales de l'Institut Henri Poincare, A20,69,(1974).
\item M.H. Goroff and A. Sagnotti, Phys.Lett.B160,81(1985), Nucl.Phys.B266,709,(1986).
\item A.E.M. van de Ven, Nucl.Phys.B378,309,(1992).
\item S. Deser and P. van Nieuwenhuizen, Phys.Rev.D10,401,(1974).
\item S. Deser, H. S. Tsao, and P. van Nieuwenhuizen, Phys.Rev.D10,3337,(1974).
\item Z. Bern, J.J. Carrasco, L.J. Dixon, H. Johansson,R.Roiban,arXiv, 1103.1848v1[hep-th].
\item Z. Bern, J.J. Carrasco, L.J. Dixon, H. Johansson, D.A. Kosower, and R. Roiban, Phys.Rev.Lett,98,161303,(2007).
\item Z. Bern, J.J. Carrasco, L.J. Dixon, H. Johansson,R.Roiban, Phys.Rev.D78,105019 (2008)[0808.4112 [hep-th]].
\item Z. Bern, J.J. Carrasco, L.J. Dixon, H. Johansson, R. Roiban, Phys.Rev.Lett. 103,081301,(2009).
\item R. Kallosh 1103.415v1[hep-th].
\item D. Gross and P. Mende, Nucl.Phys.B303,407,(1998), Phys.Lett.B197,129,(1987).
\item H. Elvang, D.Z. Freedman,and M. Kiermeir, JHEP 1010,103(2010)[1003.5018[hep-th]].
\item D.Z. Freedman and E. Tonni, 1101.1672 [hep-th].
\item S. Weinberg, Phys.Rev.B140,516,(1965).
\item T. D. Lee and M. Nauenberg, Phys.Rev.B133,1549,(1964).
\item T. Kinoshita and A. Ukawa, Phys.Rev.D13,1573,(1976).
\item I.J. Muzinich and M. Soldate, Phys.Rev.D37,359,(1988).
\item D. Amati, M. Ciafaloni, and G. Veneziano, Phys.Lett.B289,87,(1992).
\item S. Weinberg, Quantum Theory of Fields Vol 2, (Cambridge University Press, 1996)
\item F. J. Dyson, Phys.Rev.85,631,(1952).
\item L. D. Landau, Nucl.Phys. 13, 181 (1959)
\item J. D. Bjorken, Stanford University Doctoral Thesis unpublished (1959)
\item R. J. Eden, P. V. Landshoff, D. I. Olive, and J. C. Polkinghorne, The analytic S matrix (Cambridge University Press, 1966)
\end{enumerate}

\end{document}